# SPB: A Secure Private Blockchain-based Solution for Energy Trading

Ali Dorri, Fengji Luo, Salil S. Kanhere, Raja Jurdak, Zhao Yang Dong

*Blockchain is increasingly being used to provide a distributed, secure, trusted, and private framework for energy trading in smart grids. However, existing solutions suffer from lack of privacy, processing and packet overheads, and reliance on Trusted Third Parties (TTP). To address these challenges, we propose a Secure Private Blockchain-based (SPB) framework. SPB enables the energy producers and consumers to directly negotiate the energy price. To reduce the associated packet overhead, we propose a routing method which routes packets based on the destination Public Key (PK). SPB eliminates the need for TTP by introducing atomic meta-transactions. The two transactions that form a meta-transaction are visible to the blockchain participants only after both of them are generated. Thus, if one of the participants does not commit to its tasks in a pre-defined time, then the energy trade expires and the corresponding transaction is treated as invalid. The smart meter of the consumer confirms receipt of energy by generating an Energy Receipt Confirmation (ERC). To verify that the ERC is generated by a genuine smart meter, SPB supports authentication of anonymous smart meters which in turn enhances the privacy of the meter owner. Qualitative security analysis shows the resilience of SPB against a range of attacks.*

## Introduction

The grand challenges of climate change, depletion of fossil fuels, and ever-increasing energy demand are driving the transformation of energy systems. In the early 21$^{st}$ century, the concept of "smart grid" was proposed [1], with the aim of increasing efficiency, reliability and robustness.

The power distribution network which is central to smart grids is significantly characterized by high penetration of distributed renewable resources, flexible loads, and advanced sensing infrastructures. The transformation from a centralized to distributed energy generation pattern has led to the emergence of energy prosumers (Producers-and-Consumers) [2], who are capable of generating and consuming energy simultaneously, e.g., a building equipped with solar panels. This naturally raises the need for establishing an energy trading mechanism that is secure, maintains participant privacy, and fosters energy economics in the power distribution network.

Recently blockchain has attracted tremendous attention as a means to provide a secure and anonymous framework for energy trading. Blockchain employs changeable Public Keys (PKs) as the user identity to provide anonymity. In our recent work [4], a blockchain based distributed data storage system is proposed for protecting the grid data. Mihaylov *et al*. [5] propose to convert energy to a virtual currency known as NRGcoin which is then traded in a blockchain. A centralized Distribution System Operator (DSO) monitors the demand and load by collecting information from the smart meters of producers and consumers and defines the NRGcoin price accordingly. In [6], the authors implemented a blockchain information system for microgrid called "the Brooklyn microgrid" that demonstrates the potential of blockchain to underpin a distributed energy trading market. The proposed method in [7] comprises a central repository that contains all energy bids and requests in the energy market. To ensure the privacy of the energy producers, a mixing service is employed that gives a random ID to the produced energy by each smart meter. Powerledger [8] proposed a blockchain-based energy market that requires users to buy powerledger tokens to be able to trade energy. However, the following issues have yet to be addressed by the state-of-the-art:

**1) Lack of privacy:** All transactions of a user are publicly available; thus, critical information about the user such as energy consumption or production patterns can be obtained by linking multiple identities to the user or by examining the pattern of transactions in the same ledger. In most existing works, e.g., [6-7], the transactions generated by each energy prosumer can be tracked as they are either generated using the same PK or are linked together in the same ledger.

**2) Reliance on Trusted Third Party (TTP) Brokers:** Trading energy requires both sides of the transaction to fulfil their commitments in the transaction, e.g., the producer must send the consumer the energy upon receipt of payment. Achieving this level of trust is challenging due to the distributed nature of the blockchain. To address this challenge most existing methods rely on a TTP [5-8] and are therefore partially distributed. The TTP is susceptible to typical issues arising from centralization including a single point of failure, bottlenecks, etc. Additionally, the privacy of the participating entities may be compromised by the TTP as it has a complete view of all actions performed in the network. In [13] the authors proposed a new concept known as *atomic swap* which enables exchanging different cryptocurrencies in multiple blockchains without requiring exchange servers (TTP). Instead, both buyer and seller pay coin (in their respective blockchain) to a smart contract. Following, the contract transfers the exchanged coin to the buyer's account. However, this method incurs processing overhead and delay in coin exchange as four transactions must be mined, i.e., stored, in the blockchain for one exchange. A similar method is used in [9] for trading goods. Assume that Alice wishes to purchase a book from Bob. Alice creates a smart contract that serves as the TTP. Bob then pays agreed price to the contract. The contract maintains the received coin till Bob confirms receipt of the book. Although this method enables trading goods without a TTP, each trade requires at least three transactions: i) the smart contract, ii) the transaction generated by the buyer to pay to the smart contract, and iii) the transaction generated by the smart contract to pay either the seller, when the trade is successful, or the buyer, in case of dispute. This incurs processing and packet overheads in the network and increases the delay for trading goods, thus may not be scalable for large-scale smart grid networks of the future.

**3) Blockchain overhead:** Despite its benefits, blockchain consumes significant amount of computational, energy, and bandwidth resources as: i) appending (i.e., mining) a new block involves solving a resource consuming puzzle (e.g., proof of work (POW)), ii) all communications are broadcast to all nodes which increases overheads for direct negotiation between nodes. Ethereum employs Whisper [10] routing protocol that enables direct communications between multiple parties by adding destination and source fields to the transactions. However, Whisper broadcasts the transaction with only the destination node accepting it.

The main contribution of this paper is to propose a Secure Private Blockchain-based platform (SPB) to address the aforementioned challenges. Our solution can be used by pure consumers, pure producers, and prosumers. The key novel features of SPB are:

i) A routing method on top of the blockchain,
ii) A pure distributed trading method by introducing atomic meta-transactions, and
iii) A private authentication method to verify smart meters.

We first propose a new private negotiation framework that can be used by producers and consumers to negotiate the price of energy. To reduce communication overheads of negotiation compared to conventional blockchains, SPB proposes a new private routing algorithm that enables direct messaging in blockchain. SPB runs over a public blockchain, where anyone can join and participate in blockchain, and does not require any participating node to buy assets or deposit money to trade energy. SPB ensures security in the energy trading process using "atomic meta-transactions". Atomic meta-transactions are an adaptation of atomic swaps for energy trading which improves performance by reducing the associated processing overhead and delay. In an atomic meta-transaction, a constituent transaction is considered to be valid if and only if it is coupled with at least one other transaction. Once a price is agreed on, the consumer generates and broadcasts a Commit To Pay (CTP) transaction, committing to pay a specific amount of money to the producer. Unlike the state-of-the-art approaches, the consumer does not pay to the smart contract. Instead CTP freezes a specific amount of money in the customer account till the producer transfers energy to the consumer. Once energy is transferred, the consumer's smart meter generates an Energy Receipt Confirmation (ERC) transaction. The producer is paid by the smart contract after the ERC is verified. The blockchain participants must be able to ensure that an ERC transaction is signed by a genuine smart meter. Additionally, the generated ERC must remain unlinked to the previous transactions, otherwise the transactions can be tracked which may expose privacy-sensitive information about the consumer. To address this challenge, SPB introduces a Certificate of Existence (CoE). Each smart meter constructs a Merkle tree of a number of PKs and sends the root of the tree to a smart meter to sign which serves as the CoE. To use the CoE, the smart meter must sign transactions with the private key corresponding to one of the PKs in the Merkle tree. Our platform is fully decentralized whereby both entities involved in the trade will receive their benefit (money or electricity) once they perform their respective task, without relying on a TTP.

**SPB: Secure Private Blockchain-based Platform**

In SPB, the participating nodes in the smart grid including energy producers, consumers, prosumers, and distribution companies manage the blockchain by storing and verifying transactions and blocks. We assume that smart meters are tamper-resistant. To reduce the blockchain processing overhead, a lightweight consensus algorithm, e.g., Distributed Time-based Consensus (DTC) proposed in our previous work [11], is used. In DTC, each miner waits for a random time prior to mining a block, instead of solving a computationally demanding puzzle. Moreover, each miner can only mine one block within a designated *consensus_period*.

The ledger of transactions created by a producer serves as its *energy account*. To establish an energy account, the producer must create a genesis transaction. The genesis transaction can be generated by [11]: i) burning specific amount of coin, e.g., Bitcoin, meaning paying specific amount of money to an unknown address, or ii) receiving a certificate from trusted entities such as energy distributors. This prevents malicious nodes from creating fake energy accounts intending to flood the network with fake energy trading requests. Users can employ multiple energy accounts to protect against malicious nodes which may analyze the pattern of transactions in a ledger to obtain privacy-sensitive information about users which in turn may lead to user de-anonymization. Although generating multiple energy accounts increases the user anonymity and thus ensures a level of privacy, it requires the user to pay for either each additional genesis transaction or certificate. Furthermore, the user would be burdened with managing the keys used for multiple energy accounts. Thus, there is a tradeoff between increasing privacy and cost and overheads.

Each network participant can add energy to his energy account by generating a *supply_energy* transaction, which includes:

"T_ID || P_T_ID || energy_amount || energy_price || negotiatable || PK || sign"

T_ID is the unique identifier of the transaction, which is the hash of the transaction. P_T_ID is the identifier of the previous transaction generated by the same node and ensures that the user knows the public/private keys associated with an energy account, and thus has once generated a genesis transaction. The next two fields specify the amount and price of the energy. The *negotiatable* flag statues identifies whether the price is negotiable. Finally, the PK and signature of the user are populated in the transaction. The *supply_energy* transaction is sent to a smart contract which adds energy to the account of energy producer.

Our framework consists of two main phases namely: 1) Negotiation: which enables negotiation of the price of the energy, and 2) Energy Trading: which enables private energy trading. The two phases are explained below.

### Phase1) Negotiation

The negotiation is conducted using the *negotiation* transaction, which includes:

"T_ID || energy_account.PK || price || status"

where T_ID is the identity of the transaction, *energy_account.PK* is the PK of the energy account with which the consumer wishes to negotiate. The *price* field is the offered price by the consumer. The *status* field can be either '1' or '0', indicating that the offer is accepted or rejected by the producer, respectively. A producer that receives the negotiation transaction, can accept the request by setting status to '1' or offer another price (status set to '0'). Thus, the producer and consumer may directly exchange negotiation transactions till they agree on the price. The negotiation transactions are not stored in the blockchain as they form intermediate communications between the two parties. The final negotiated price is later included in energy trading transactions as outlined in Phase 2.

Recall that in conventional energy trading approaches, transactions are typically broadcast which incurs packet overheads and delays for processing negotiation between nodes. To address this challenge, we propose an *Anonymous Routing Backbone (ARB)*. The participating nodes in blockchain which have sufficient resources, e.g., substations or controller centers, serve as the backbone nodes which are responsible for routing negotiation transactions to the destination. Unlike traditional networks where participating nodes are known by their Internet Protocol (IP) address, blockchain participants are known by their PK. Thus, ARB is based on the PK of the nodes, meaning that packets are routed toward specific PKs. The backbone nodes route transactions based on the *X* most significant bytes of the PK of the destination, known as Routing Bytes (RB). The backbone nodes initialize a Distributed Hash Table (DHT), such as the example in Figure 1, that identifies the corresponding backbone node to each value of RB. The value of X is influenced by the total number of backbone nodes and the total load, i.e., transactions that need to be routed. In small backbones, X is very small (e.g., X=1 in Figure 1), while for larger backbone X increases. If a backbone node is overloaded by transactions, then the backbone nodes reconstruct the hash table and update RB to a larger value. Thus, the number of transactions that arrive at each backbone node, i.e., match with RB of each backbone node, will reduce. It is worth noting that although increasing the RB reduces the load, it does not guarantee balancing the load between backbone nodes as the most significant bytes in PK are randomly generated. Participating nodes in the blockchain associate with the backbone nodes based on their PK which is detailed in the rest of this section. The backbone nodes use conventional routing protocols to route transactions toward destination, e.g., OSPF [RFC2328].

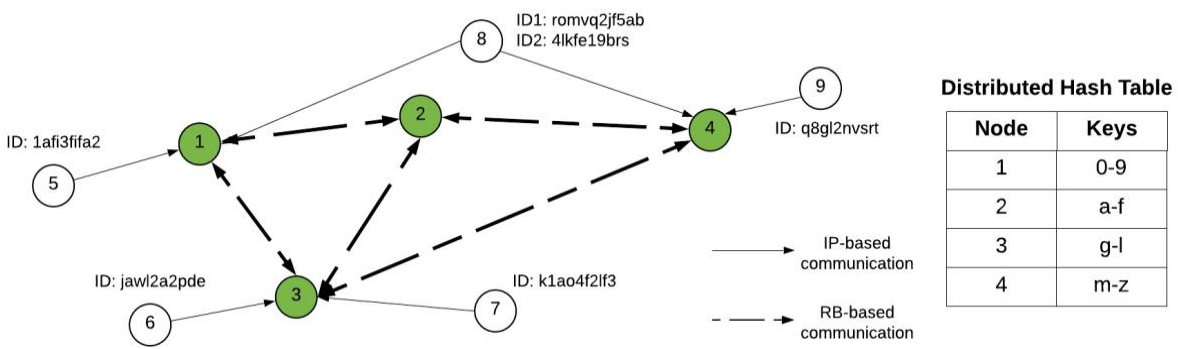

Figure 1. An illustrative example of ARB and the corresponding hash table.

When the backbone is formed, all nodes in the blockchain, referred to as regular nodes in the rest of the paper, are notified of the DHT. Each regular node then associates with the backbone node that handles transactions for its PK. Since a regular node may employ multiple PKs, it may be associated with multiple backbone nodes, e.g. node 8 in Figure 1. For each of its PKs, the regular node joins the backbone node by sending a join message to the backbone node, which is signed with the private key corresponding to the PK. This protects against malicious nodes that falsely claim to be another node as only the genuine node knows the private key for signing a message. The backbone nodes are responsible for forwarding transactions to the regular nodes that are associated with them. Regular nodes use the IP address of the backbone nodes to send transactions as shown in Figure 1. However, the backbone nodes use the destination PK of the received transaction to determine the destination node and the corresponding IP address, to decide on which path this transaction must be forwarded along. Compared to traditional IP-based routing, ARB incurs a small processing overhead at the backbone nodes for looking up the destination PK. To protect against malicious nodes that may monitor the IP address of transactions to deanonymize a user, mechanisms such as Tor can be used for anonymization between regular and ARB nodes.

To negotiate with a producer, the consumer sends an offer with its suggested price and energy requirement. The negotiation message is sent to one of the backbone nodes and is subsequently routed to the producer. The response message from the producer follows the same process but in the reverse direction towards the consumer. The maximum number of offers and counteroffers that can be generated is limited to a certain threshold, *offer_limit*. The latter is a value defined by the system designer based on the application that prevents malicious nodes from launching a denial of service attack by sending a large number of negotiation messages.

*Phase2) Energy Trading*

After negotiations (if any), the consumer and the producer commence the energy trading process. We eliminate the need for a TTP, which is common to conventional energy trading methods, by introducing *atomic meta-transactions* based on atomic swaps [12]. An atomic process is an indivisible operation which appears to the rest of the system to occur at once without being interrupted. We define two transactions that together form atomic meta-transactions, such that they are valid if and only if both transactions are generated within a specific time period. A single transaction is not valid by itself.

The atomic meta-transaction process is outlined in Figure 2. The first transaction is generated by the consumer to commit payment of the price agreed with the producer, and thus is known as *Commit To Pay (CTP)* transaction. The CTP is a payment transaction that pays a specific amount of money to another user, which is the producer. However, CTP payment has *pending* status meaning that the consumer cannot spend the money committed in CTP, because the money has not yet been transferred to the producer's account. The CTP is not mined in the blockchain. Each miner maintains a database of pending CTPs, which is used for verification. To ensure the consistency of the CTP database among blockchain participants, a new header field is introduced for blocks named "*CTP hash*". The latter is populated by each miner with the hash of its CTP database copy. The issue of maintaining consistency among the CTP databases at the miners is similar to ensuring that the local blockchain copies at these nodes are consistent. The CTP database reduces delay and processing overheads for trading energy compared to conventional blockchain-based solutions as the CTP transaction does not need to be mined. To secure SPB against double spending, where a malicious node pays the same amount of coin to multiple users, we propose an approach for transaction verification which is different to conventional blockchains. In the latter, the miners verify a payment transaction by ensuring that the amount of money being spent is less than or equal to the amount of money that the user has in his account. However, in the case of SPB, the miners check if the amount of money to be spent is less than the sum of the amount of money in the user's account in blockchain and the CTP database.

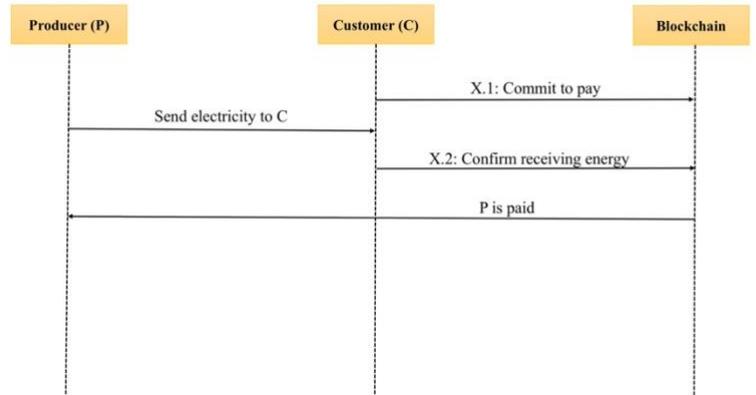

Figure 2. The process of atomic meta-transactions.

The structure of CTP transaction is as follows:

*"T_ID || Time Stamp || Expiry Time|| Price || Contract Hash || PK || Sign"*

*T_ID* is the transaction identifier. *Time Stamp* is the time when the transaction is generated. Each CTP transaction can only be stored by the miners for a specific period of time, which is indicated in the *Expiry Time* field. This ensures that if the producer does not send energy to the consumer, the pending money of the consumer will be returned to his account after the expiry time. *Price* denotes the money that needs to be transferred between producer and the consumer. *Contract Hash* is the hash of the amount and rate of energy that the producer and consumer have agreed on (either by negotiation or directly). The final two fields are the PK and signature of the consumer. The consumer signs the hash of the entire transaction to ensure integrity.

Once the CTP is broadcast, the producer (as a member of the network) receives the transaction and compares the contract hash with its own contract hash, which is stored locally, to ensure the consumer has not changed the amount or rate of the energy. To ensure privacy of both producer and consumer, rather than including the exact amount and rate of energy, a hash is used. Next, the producer begins delivering energy to the smart meter of the consumer. Once all of the energy is delivered, the consumer's smart meter generates an *Energy Receipt Confirmation (ERC)* transaction. The network participants need to verify that a node that claims to be a meter is genuine to protect against malicious activities. Conversely, the meter must remain private as it reveals sensitive information including electricity usage of its owner. To address this challenge, we propose a method to anonymously verify a smart meter without revealing the true identity of the meter using a Certificate of Existence (CoE). Figure 3 provides a summary of CoE generation and verification process which are elaborated below. The manufacturer of each meter creates a private/public key pair for each smart meter (known as M-PK$^-$/*M-PK$^+$* respectively) and serves as the Certified Authority (CA) for the M-PK$^+$ of all the meters. Once the smart meters are deployed, each meter produces a number of public/private keys (step (1)). The smart meter owner determines the number of keys required depending on the level of anonymity desired. A large pool of key pairs can achieve higher anonymity, as transactions which are generated using the same key can be linked thus revealing energy usage patterns. Instead when multiple keys are used, the energy usage history of the user is split in different ledgers which have no mutual links, thus making the aforementioned inferences difficult. After generating the key pairs, the smart meter constructs a Merkle tree (2) by recursively hashing the PKs which are stored as the leaves of a tree as shown in Figure 4. The PKs in the Merkle tree and their corresponding private keys are later used to privately create ERC transaction, as outlined in the rest of this section. A key feature of Merkle tree is that the existence of a leaf can be proved with small overhead. As an illustrative example, to prove the existence of 'A' in the Merkle tree shown in Figure 4, one must store $H_B$ and $H_{CD}$ locally and reveal them to prove existence. To verify the existence, $H_A$ is hashed by $H_B$, and the result is then hashed by $H_{CD}$. If the final hash matches with $H_{ABCD}$, then the existence of 'A' is confirmed. The meter generates a Verification Request (VR) with the following fields:

$$Root \;||\; M\text{-}PK^+ \;||\; Sign$$

Where *Root* is root hash of the Merkle tree encrypted with the M-PK$^+$ of a randomly chosen smart meter, known as Verifier Meter (VM). The second field is the M-PK$^+$ of the meter that generates VR, followed by Sign that is the signed hash of the VR by *M-PK$^-$*. Once populated, the meter routes VR to the VM using the proposed routing method (3). On receiving the Merkle tree root, the VM first verifies the smart meter, that has generated the root, using its M-PK$^+$ and the CA of the manufacturer (4). If verified, it signs the received root using its M-PK⁻ (5) and returns the signed root back to the smart meter of which initially sent the transaction. The signed root of the Merkle tree is used as the CoE.

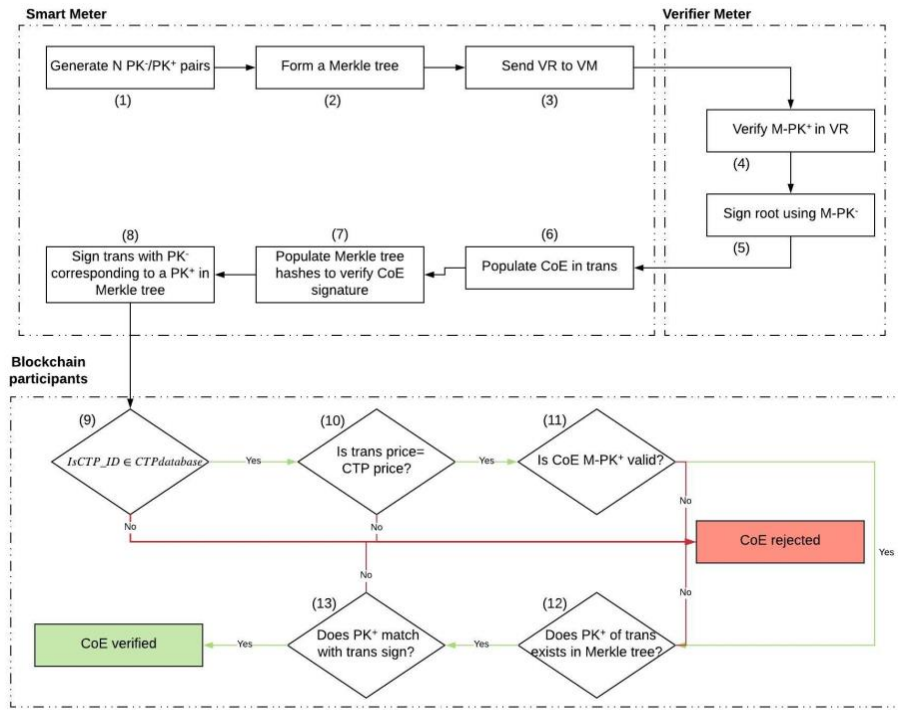

Figure 3. A summary of CoE generation and usage process.

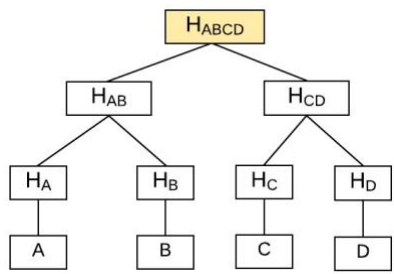

Figure 4. The structure of a Merkle tree.

In order to generate the ERC transaction, the smart meter adopts one of the keys used to construct the Merkle tree in CoE as the identity of the ERC transaction (6). The structure of ERC transaction is as follows:

*"T_ID || Time Stamp || CTP_ID || Price || CoE || CoE_PK || Merkle hashes || PK|| Sign"*

The first two fields are the same as the CTP transaction. The *CTP_ID* refers to the ID of the corresponding CTP transaction that is used by the miners to verify the ERC transaction as discussed below. *Price* denotes the total price of the contract. The CoE is as discussed above. The CoE_PK contains the PK of the VM used for verification of the CoE using CA. Merkle hashes are the hashes that are required to prove the existence of the PK used to generate the ERC transaction in the Merkle tree (7). Finally, the last two fields are the PK of the ERC generator and its corresponding signature (8).

On receiving the ERC transaction, the miners verify the transaction by: a) verifying that the *CTP_ID* exists in the list of pending transactions (9), b) matching the price of the current transaction with the CTP price (10), c) verifying the signature and PK of the VM using the manufacturer CA (11), d) verifying the existence of PK used to generate the transaction in the Merkle tree (12), and e) matching the PK hash in the Merkle tree with the signature on the transaction (13). If the above steps are successfully performed, then the ERC is validated. The ERC triggers the smart contract to pay the agreed price to the producer. The smart contract cannot be triggered if even one of the steps is not successful and the CTP transaction of the consumer is removed from the CTP database after the expiry time.

To demonstrate the use of SPB by end users, we implemented the core SPB functions using smart contracts in Ethereum. A video demonstration is available at [14]. Since we have used Ethereum for this demonstration, optimization methods employed by SPB are not accommodated, e.g., CTP is stored in the blockchain instead of a CTP database.

## Security and Privacy Analysis

In this section, we discuss the security and privacy of SPB.

*Privacy:* We study the privacy of SPB from the perspective of the producer, consumer, and the consumer's smart meter. The producer can employ multiple energy accounts which protect him from being tracked. The consumer and his smart meter employ changeable PKs for communicating with multiple users to enhance his anonymity and thus provide a level of privacy.

Recall that CoE is proposed to protect the privacy of the consumer. There is no link between the VM and the actual smart meter that is using the CoE as VM is selected randomly by the smart meter. Each meter may sign multiple CoEs for other smart meters. Consequently, tracking the CoEs signed by particular meters will not compromise the privacy. The smart meter changes the PK used for each ERC transaction, thus, protecting against potential tracking by malicious nodes.

The Merkle tree in a CoE may be used by a single meter or by multiple meters that protect against malicious nodes who track a particular CoE.

*Security:* The security of SPB draws on the inherent security features of blockchain. Each transaction contains the hash of either the contract or the whole transaction that ensures integrity. We discuss below the key attacks possible in SPB:

**Malicious producer:** A malicious producer may not deliver energy to the consumer after the consumer generates CTP. In this case, the smart meter does not generate the corresponding ERC transaction as energy is not received. The CTP transaction will be discarded from CTP database by the expiry time and the consumer's money will be released.

**Malicious consumer:** The malicious consumer cannot receive energy without paying the producer as the energy transfer can only be triggered by a valid CTP. Recall that it is assumed that smart meters are tamper-resistant and thus it is impossible to generate a fake ERC transaction.

**CoE attack:** The malicious node pretends to be a smart meter by using the CoE of a genuine smart meter. ERC transaction is signed by one of the keys in the Merkle tree. Thus, the malicious node must possess the private key corresponding to one of the keys in the Merkle tree which are only known by genuine smart meters.

## Conclusion

We proposed a Secure Private Blockchain-based framework (SPB), that enables energy prosumers to negotiate over the energy price and trade energy in a distributed manner. SPB employed a routing method for forwarding negotiation traffic, thus reducing the associated packet overhead compared to existing approaches. SPB eliminated the need for TTP using atomic meta-transactions. Smart meters are verified using Certificate of Existence (CoE) without revealing information about the previous transactions of the same meter which in turn enhanced user privacy. Security analysis showed the robustness of SPB against several attacks. In the future, we aim to implement SPB on real devices and benchmark its performance; we are also working on developing more Blockchain based applications in the smart grid domain.

## Biographies

Ali Dorri received his bachelor degree in Computer Engineering from Bojnourd University, IRAN, 2012. He then commenced his master degree in Computer Engineering in Islamic Azad University of Mashhad, IRAN, working on Mobile Ad hoc Networks and security issues rising from this sort of network. He now is a Ph.D. candidate in University of New South Wales (UNSW), Sydney. His current research interest covers security and privacy concerns in the context of Internet of Things (IoT), Wireless Sensor Network (WSN) and Vehicular Ad hoc Network (VANET). Moreover, he is working on bitcoin and its applications on IoT.

Fengji Luo received his bachelor and master degrees in Software Engineering from Chongqing University, China in 2006 and 2009. He received his Ph.D degree in Electrical Engineering from The University of Newcastle, Australia, 2014. Currently, he is a Lecturer and Academic Fellow in the School of Civil Engineering, The University of Sydney, Australia. His research interests include energy demand side management, smart grid, and energy informatics. He has published over 100 papers on peer referred journals and conferences. He receives the Pro-Vice Chancellor's Research and Innovation Excellence Award of The University of Newcastle in 2015 and the UUKI Rutherford Fellowship in 2018.

Salil S. Kanhere received his M.S. and Ph.D. degrees, both in Electrical Engineering from Drexel University, Philadelphia. He is currently an Associate Professor in the School of Computer Science and Engineering at the University of New South Wales in Sydney, Australia. His research interests include Internet of Things, pervasive computing, crowdsourcing, blockchains, mobile networking, privacy and security. He has published over 185 peer-reviewed articles and delivered over 20 tutorials and keynote talks on these research topics. He is a contributing research staff at Data61, CSIRO and a faculty associate at Institute for Infocomm Research, Singapore. Salil regularly serves on the organizing committee of a number of IEEE and ACM international conferences (e.g., IEEE PerCom, ACM MobiSys, ACM SenSys, ACM CoNext, IEEE WoWMoM, IEEE LCN, ACM MSWiM, IEEE DCOSS,). He currently serves as the Area Editor for Pervasive and Mobile Computing, and Computer Communications. Salil is a Senior Member of both the IEEE and the ACM. He is a recipient of the Humboldt Research Fellowship in 2014.

Raja Jurdak is a Senior Principal Research Scientist at CSIRO, where he leads the Distributed Sensing Systems Group. He has an MSc in Computer Engineering and a PhD in Information and Computer Science, both from the University of California. His current research interests focus on energy-efficiency and mobility in networks. He has over 130 peer-reviewed journal and conference publications, as well as a book published by Springer in 2007 titled Wireless Ad Hoc and Sensor Networks: A Cross-Layer Design Perspective. He regularly serves on the organizing and technical program committees of international conferences (DCOSS, RTSS, PERCOM, EWSN, ICDCS, Blockchain, MSWIM, WoWMoM). He is a Guest Editor for a Special Issue of Elsevier IoT Journal on Distributed Ledger Technology (DLT) for the Internet of Things. Raja is an Honorary Professor at University of Queensland, and an Adjunct Professor at Macquarie University, James Cook University, and the University of New South Wales. He is a Senior Member of the IEEE.

Z.Y. Dong obtained Ph.D. from the University of Sydney, Australia in 1999. He is now with the University of NSW, Sydney, Australia. His immediate role is Professor and Head of the School of Electrical and Information Engineering, The University of Sydney. He also served as director for Faculty of Engineering and IT Research Cluster on Clean Intelligent Energy Networks, and the Sydney Energy Internet Research Institute. He was Ausgrid Chair and Director of the Ausgrid Centre for Intelligent Electricity Networks (CIEN) providing R&D support for the $600M Smart Grid, Smart City national demonstration project. He also worked as manager for (transmission) system planning at Transend Networks (now TASNetworks), Australia. His research interest includes smart grid, power system planning, power system security, load modeling, renewable energy systems, and electricity market. He is an editor of IEEE Transactions on Smart Grid, IEEE Transactions on Sustainable Energy, IEEE PES Transaction Letters and IET Renewable Power Generation. He is an international Advisor for the journal of Automation of Electric Power Systems. He is Fellow of IEEE.